\documentclass[trackchanges,twocolumn]{aastex7}
\usepackage{mathrsfs}
\usepackage{amsmath}

\begin{document}

\title{Disk Instability Model for Quasi-Periodic Eruptions: Investigating Period Dispersion and Peak Temperature}

\author[orcid=0000-0002-6938-3594]{Xin Pan}
\affiliation{National Astronomical Observatories, Chinese Academy of Sciences, 20A Datun Road, Beijing 100101, People’s Republic of China}
\email[show]{panxin@bao.ac.cn}  

\author[orcid=0000-0002-7299-4513]{Shuang-Liang Li}
\affiliation{Key Laboratory for Research in Galaxies and Cosmology, Shanghai Astronomical Observatory, Chinese Academy of Sciences, 80 Nandan Road, Shanghai 200030, People’s Republic of China}
\email[show]{lisl@shao.ac.cn}

\author[orcid=0000-0002-2355-3498]{Xinwu Cao}
\affiliation{Institute for Astronomy, School of Physics, Zhejiang University, 866 Yuhangtang Road, Hangzhou 310058, People’s Republic of China}
\email[show]{xwcao@zju.edu.cn}

\author[orcid=0000-0001-9920-4019]{Bifang Liu}
\affiliation{National Astronomical Observatories, Chinese Academy of Sciences, 20A Datun Road, Beijing 100101, People’s Republic of China}
\affiliation{School of Astronomy and Space Science, University of Chinese Academy of Sciences, 19A Yuquan Road, Beijing 100049, People’s Republic of China}
\email[]{}

\author[]{Weimin Yuan}
\affiliation{National Astronomical Observatories, Chinese Academy of Sciences, 20A Datun Road, Beijing 100101, People’s Republic of China}
\affiliation{School of Astronomy and Space Science, University of Chinese Academy of Sciences, 19A Yuquan Road, Beijing 100049, People’s Republic of China}
\email[]{}

\correspondingauthor{Xin Pan, Shuang-Liang Li, Xinwu Cao}


\begin{abstract}
Quasi-periodic eruptions (QPEs) are a class of X-ray repeating burst phenomena discovered in recent years. Many models have been proposed to study this phenomenon, there remains significant debate regarding the physical origin of QPEs. In our previous work, we developed a disk instability model with a large-scale magnetic field and successfully reproduced the light curves and spectral characteristics of several QPE sources. We further investigate the model in this work, aiming to explain two key observational features: the dispersion in eruption periods and the peak temperatures during eruptions. The model reveals critical thresholds ($\dot{M}_{\rm crit}$, $\beta_{1,\rm crit}$) that separate systems into stable regimes with minimal period variations and unstable regimes where periods are highly sensitive to accretion rate and magnetic field parameter, while peak temperatures remain nearly constant across the parameter space. This framework successfully explains both the regular eruptions observed in sources like GSN 069 and the stochastic behavior in sources like eRO-QPE1, and simultaneously accounting for the observed temperature stability during long-term QPEs evolution.

\end{abstract}



\section{Introduction} 
A new type of extreme X-ray transient, known as QPEs, was first identified in 2019 through their characteristic repeating bursts from the low-mass supermassive black hole (SMBH) in GSN 069 \citep{2019Natur.573..381M}. These events exhibit recurrence timescales ranging from hours to days and are typically associated with SMBHs of $10^5$ to several $10^7$ solar masses. Their X-ray spectra are well described by two components: a stable disk blackbody and a variable thermal component with characteristic temperatures of 0.1-0.2 keV. To date, about 10 sources have been reported as confirmed or candidate QPEs \citep{2019Natur.573..381M,2020A&A...636L...2G,2021Natur.592..704A,2021ApJ...921L..40C,2023A&A...675A.152Q, 2024NatAs...8..347G,2024A&A...684A..64A,2024Natur.634..804N,2025ApJ...983L..39C,2025NatAs.tmp...99H,2025arXiv250617138A}. This growing but still limited sample presents both opportunities and challenges for theoretical models.

Observational studies of QPEs have identified several defining characteristics that provide critical constraints for theoretical models, including: alternating recurrence times in some sources (e.g. GSN 069) suggesting two eruptions may form complete cycles \citep{2021ApJ...921L..32X,2023A&A...675A.100F}; a
counter-clockwise hysteresis cycle in the luminosity-temperature (L-T) plane \citep{2022A&A...662A..49A,2023A&A...670A..93M,2024A&A...684A..64A}; several QPE sources show archival evidence of prior tidal disruption events (TDEs) \citep{2023A&A...675A.152Q,2024Natur.634..804N,2025ApJ...983L..39C,2025MNRAS.540...30B}, and GSN 069 \citep{2018ApJ...857L..16S,2021ApJ...920L..25S,2023A&A...670A..93M,2025ApJ...978...10K}, XMMSL1 J024916.6-041244 \citep{2021ApJ...921L..40C}, and eRO-QPE3 \citep{2024A&A...684A..64A} also appear likely to be X-ray TDEs despite lacking optical flare detections, suggesting a possible physical connection; source-dependent variations in recurrence time dispersion \citep{2024Natur.634..804N}; stable peak temperatures ($\sim0.1-0.2$ keV) across outbursts despite significant luminosity variations \citep{2024ApJ...965...12C}. These collective properties challenge any comprehensive QPE model to simultaneously account for both the temporal and thermal behaviors observed across different sources.

Many theoretical models have been proposed to explain this newly discovered phenomenon, which generally fall into two broad categories: 1) disk instability models that better reproduce the observed spectral properties through detailed radiative processes \citep{2022ApJ...928L..18P,2023ApJ...952...32P,2023MNRAS.524.1269K}, and 2) models related to orbital motion that more accurately predict the observed timing characteristics \citep{2020MNRAS.493L.120K,2021MNRAS.503.1703I,2023A&A...675A.100F,2024PhRvD.109j3031Z,2025A&A...693A.179M}. These models demonstrate good agreement with observations for sources exhibiting relatively regular eruption patterns, such as GSN 069 and eRO-QPE2. However, both theoretical frameworks demonstrate significant limitations when addressing more complex observational behaviors, such as the highly dispersed recurrence times in eRO-QPE1 \citep{2024ApJ...965...12C} or the consistently stable peak temperatures across eruptions with varying amplitudes in certain QPE sources \citep{2025arXiv250412762G}. 

Orbital motion models attribute the quasi-periodic nature of QPEs to a companion star passing through specific orbital positions, with proposed mechanisms including self-lensing effects, peri-center passage interactions, or direct star-disk \citep{2020MNRAS.493L.120K,2021MNRAS.503.1703I,2024PhRvD.110h3019Z} collisions (also possible black hole (BH)-disk and stream-disk collisions, \citealt{2023A&A...675A.100F,2025arXiv250510611Y,2025arXiv250610096L}). These models predict highly stable recurrence times when the orbital parameters and accretion structure remain fixed \citep{2025A&A...693A.179M}. However, current implementations lack self-consistent treatment of the radiative processes during eruptions. Although recent studies have investigated temperature predictions from star-disk collisions \citep{2023A&A...675A.100F,2023MNRAS.526...69T,2025arXiv250412762G,2025ApJ...978...91Y}, their results remain inconsistent with the narrow observed temperature range of $\sim0.1-0.2$ keV across QPE sources. In contrast, disk instability models naturally explain the observed eruption temperatures through radiative processes in the inner accretion flow \citep{2022ApJ...928L..18P,2023ApJ...952...32P}. A question is the sensitivity of model to disk perturbations, a key factor in explaining irregular QPEs like eRO-QPE1, which has not been systematically examined. By analyzing how both the periodicity and thermal properties in disk instability model respond to accretion disk perturbations, we can evaluate whether this framework can unify both regular and stochastic QPE populations within a single physical scenario.

In this work, we will investigate the disk instability model of QPEs established by \cite{2022ApJ...928L..18P}, examining the relationships between the model's eruption period/peak temperature and local accretion rate/magnetic field parameter. Our analysis aims to assess whether this model can adequately explain both the observed period dispersion and the peak temperatures evolution across different QPEs.

\section{Model}
Standard accretion disks at moderate accretion rates ($\sim0.01-0.5$ Eddington accretion rate) are susceptible to thermal and secular instabilities in their radiation pressure dominated inner regions \citep{1973A&A....24..337S,1976MNRAS.175..613S}. This well established instability mechanism successfully explains recurrent variability in X-ray binaries \citep{2016ApJ...833...79W}, and has been suggested to possibly apply also to active galactic nuclei (AGN, \citealt{2020A&A...641A.167S}). However, observed variability timescales in some AGNs, such as Changing-look AGNs (CL AGNs) and extremely variable quasars (EVQs), often deviate from standard viscous timescale predictions, presenting the so-called 'viscosity crisis' \citep{2018NatAs...2..102L}. Recent studies have demonstrated that incorporating magnetic fields into accretion disk models can alleviate this discrepancy, thereby providing a possible explanation for short-timescale AGN variability phenomena \citep{2019MNRAS.483L..17D,2021ApJ...910...97P,2021ApJ...916...61F,2023A&A...672A..19S}.

Since this work mainly focus on the eruption properties of QPEs rather than complete spectral modeling of the disk, we adopt the accretion disk model of \cite{2022ApJ...928L..18P}, in which a large-scale magnetic field modifies both angular momentum transport and instability development. The key equations governing the steady-state disk structure are:
\begin{equation}
    \frac{\mathrm{d}\dot{M}}{\mathrm{d}R}+4\pi R\dot{m}_{\rm w}=0,
    \label{continuity}
\end{equation}
\begin{equation}
    -\frac{1}{2\pi}\frac{\mathrm{d}(\dot{M}l_{\rm k})}{\mathrm{d}R}-\frac{\mathrm{d}}{\mathrm{d}R}(R^{2}\mathscr{B}\mathscr{C}^{-1/2}\mathscr{D}T_{r\phi})+T_{\rm m} R=0,
    \label{angularmom}
\end{equation}
\begin{equation}
    P_{\rm tot}=(1+\frac{1}{\beta_1})(P_{\rm gas}+P_{\rm rad}),
    \label{EoS}
\end{equation}
\begin{equation}
    -\frac{3}{2}\Omega_{\rm k}T_{r\phi}\frac{\mathscr{BD}}{\mathscr{C}}=\frac{8acT_{\rm c}^{4}}{3\tau},
    \label{energy}
\end{equation}
where $\beta_{1}=(P_{\rm gas}+P_{\rm rad})/P_{\rm m}$, with $P_{\rm gas}$, $P_{\rm rad}$ and $P_{\rm m}$ denoting the gas, radiation, and magnetic pressures, respectively. To account for the observed QPE recurrence timescales, the radiation pressure instability region must be magnetically confined to an extremely narrow belt ($\sim0.1R_{\rm s}$) adjacent to the innermost stable circular orbit (ISCO). Under such extreme confinement, the temperature and surface density evolution of the unstable region can be effectively described by a one-zone model:
\begin{equation}
\begin{split}
    &\left[u^t-\frac{C_{\rm H}H\left(1-\beta_{2}\right)}{\Sigma\left(1+\beta_{2}\right)}\right]\frac{\mathrm{d}\Sigma}{\mathrm{d}t}+\frac{C_{\rm H}H\left(4-3\beta_{2}\right)}{T\left(1+\beta_{2}\right)}\frac{\mathrm{d}T}{\mathrm{d}t}\\
    &-\frac{\dot{M}_{0}-\dot{M}-4\pi R\dot{m}_{\rm w}\Delta R}{2\pi R\Delta R}=0,
    \label{sur_den_evolution}
\end{split}
\end{equation}
\begin{equation}
\begin{aligned}
    \frac{\mathrm{d} T}{\mathrm{d}t}=&\frac{T(Q^+-Q^--Q_{\rm adv})(1+\frac{1}{\beta_1})(1+\beta_2)}{2PHu^{t}(28-22.5\beta_2-1.5\beta_2^2+\frac{12-9\beta_2}{\beta_1})}\\
    &+2\frac{T\mathrm{d}\Sigma}{\Sigma\mathrm{d}t}\frac{4-3\beta_2+\frac{2-\beta_2}{\beta_1}}{28-22.5\beta_2-1.5\beta_2^2+\frac{12-9\beta_2}{\beta_1}},
    \label{temp_evolution}
\end{aligned}
\end{equation}
where $\beta_{2}=P_{\rm gas}/(P_{\rm gas}+P_{\rm rad})$, and all other parameters in Equations (\ref{continuity})-(\ref{temp_evolution}) follow exactly those presented in \cite{2022ApJ...928L..18P}.

This model incorporates seven fundamental parameters: black hole mass ($M$), accretion rate ($\dot{M}$), magnetic field parameter ($\beta_{1}$), viscosity parameter ($\alpha$), black hole spin parameter ($a$), viscous torque parameter ($\mu$), and ISCO torque condition ($f$). Generally, the $M$ and $a$ can be treated as constant over QPEs evolution timescale. Since the physical origins of $\alpha$, $\mu$, and $f$ remain incompletely understood, we just treat them as fixed parameters in our model, following our previous work \citep{2022ApJ...928L..18P}. Therefore, we focus specifically on examining how perturbations in two key local accretion parameters, $\dot{M}$ and $\beta_{1}$, affect both the periodicity and peak temperature of the repeating bursts.

\section{Results}
Our analysis begins with a fiducial parameter set: $M=10^{6}M_{\odot}$, $\dot{M}=0.1\dot{M}_{\rm Edd}$, $\beta_1=40$, $\alpha=0.1$, $a=0.98$, $\mu=0.15$, and $f=0.9$, which are commonly used in \cite{2022ApJ...928L..18P,2023ApJ...952...32P}. We systematically probe the parameter space by independently varying accretion rate $\dot{M}$ and magnetic field parameter $\beta_{1}$ to isolate their effects on eruption properties. The coupling behavior of these parameters are analyzed in the Appendix.

\subsection{repeating period vs. $\dot{M}$/$\beta_{1}$}
\begin{figure}
    \centering
    \includegraphics[width=1\linewidth]{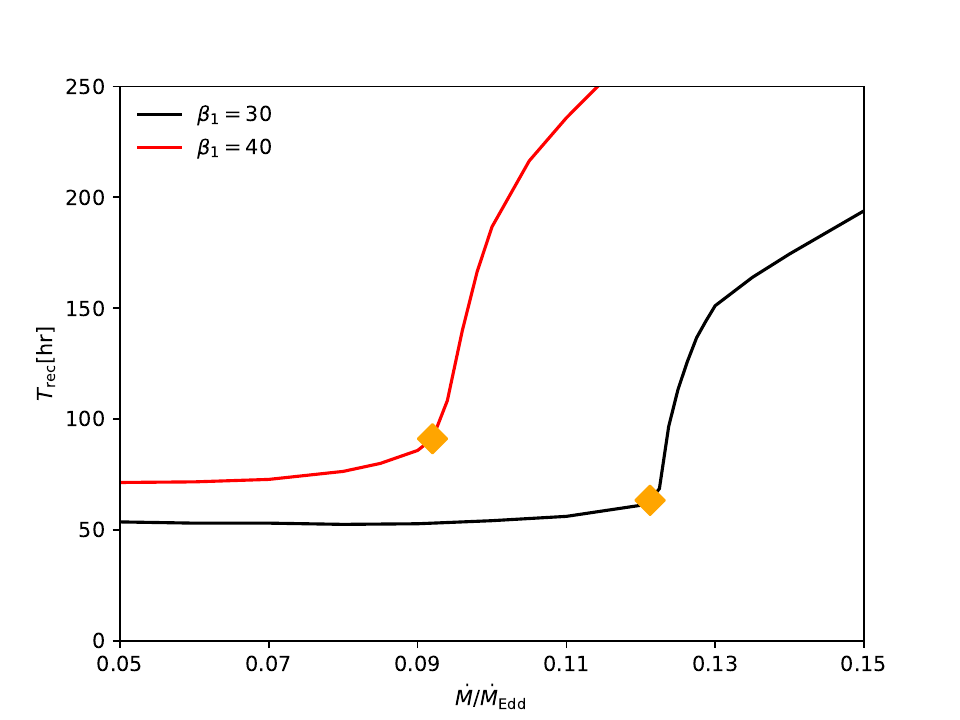}
    \caption{The Recurrence time as functions of accretion rate. The red line shows the fiducial model with $\beta_{1}=40$, and the black line corresponds to $\beta_{1}=30$. The yellow diamonds mark critical values $\dot{M}_{\rm crit}$ in this relation: when $\dot{M}<\dot{M}_{\rm crit}$ recurrence time shows weak dependence on $\dot{M}$; when $\dot{M}>\dot{M}_{\rm crit}$ recurrence time become strongly $\dot{M}$ dependent.}
    \label{fig:rec_mdot}
\end{figure}

We first examine how accretion rate variations influence eruption periodicity. Figure \ref{fig:rec_mdot} demonstrates a clear bifurcation in model behavior at a critical accretion rate $\dot{M}_{\rm crit}$. For $\dot{M}>\dot{M}_{\rm crit}$, the recurrence time $T_{\rm rec}$ becomes highly sensitive to $\dot{M}$ variations, whereas $T_{\rm rec}$ remains effectively constant when $\dot{M}<\dot{M}_{\rm crit}$. This critical transition consistently appears under different magnetic field configurations (the similar behavior at both $\beta_{1}=30$ and $\beta_{1}=40$ in Figure \ref{fig:rec_mdot}).

\begin{figure}
    \centering
    \includegraphics[width=1\linewidth]{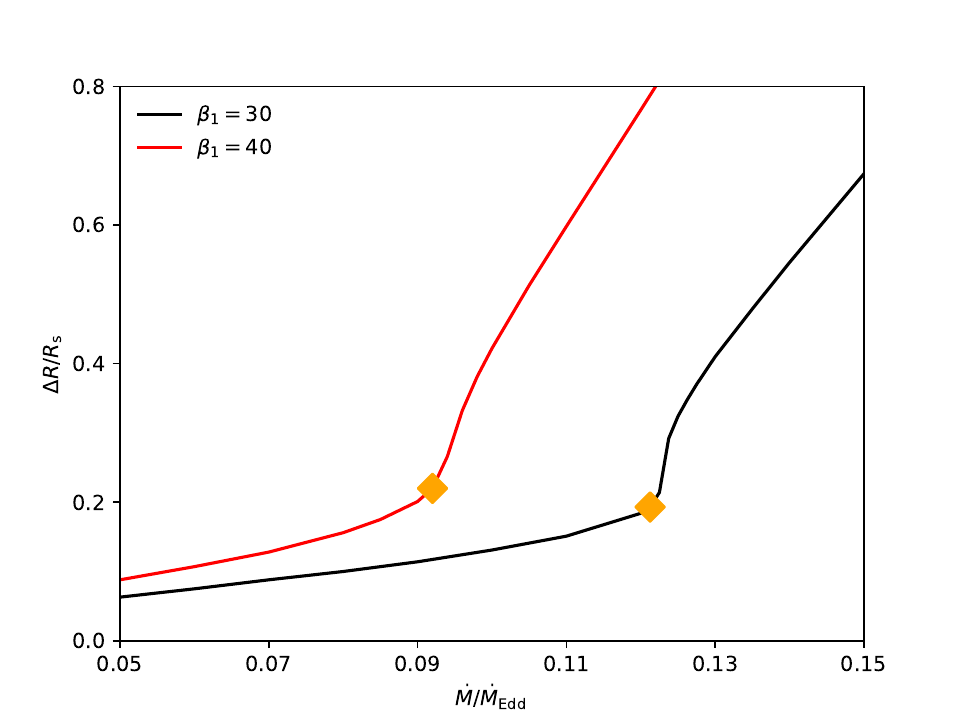}
    \caption{Similar to Figure \ref{fig:rec_mdot}, but showing the relation between the width of unstable region and accretion rate.}
    \label{fig:deltaR_mdot}
\end{figure}

\begin{figure}
    \centering
    \includegraphics[width=1\linewidth]{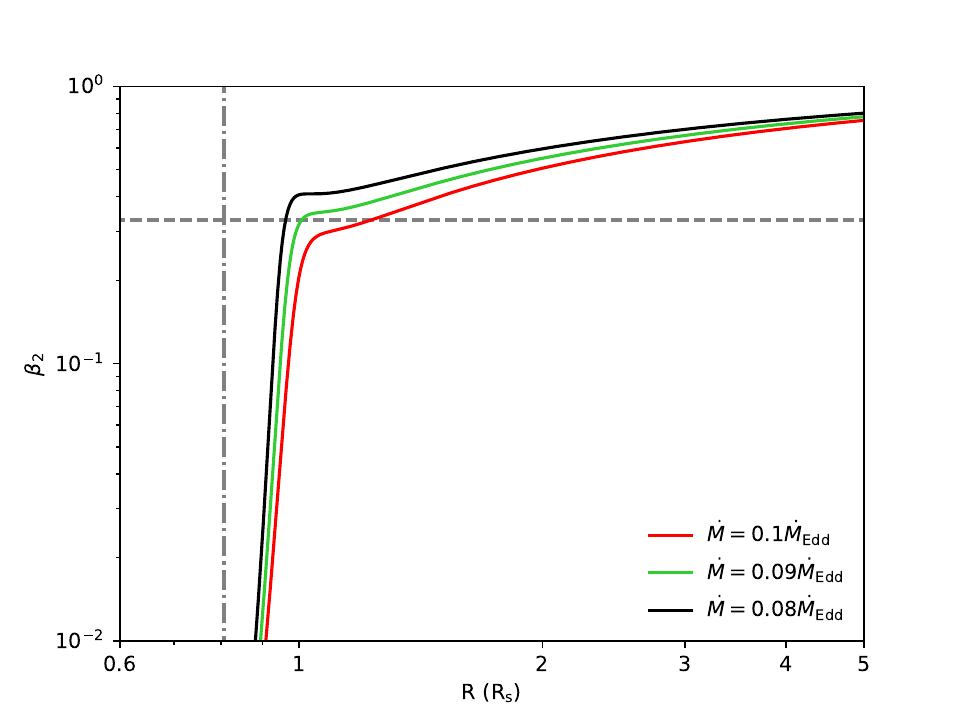}
    \caption{The ratio of gas pressure to the sum of gas and radiation pressures, $\beta_{2}$, as a function of radius. The red line shows the fiducial model ($\dot{M}=0.1\dot{M}_{\rm Edd}$), while the green and black lines represent cases with modified accretion rates: $\dot{M}=0.09\dot{M}_{\rm Edd}$ and $\dot{M}=0.08\dot{M}_{\rm Edd}$, respectively. The horizontal dashed line indicates the instability criterion ($\mu=0.15$), and the regions below this line are radiation pressure dominated and unstable. The vertical dash-dot line marks the ISCO position.}
    \label{fig:beta2_R}
\end{figure}

The sharp transition in recurrence time $T_{\rm rec}$ at the critical accretion rate $\dot{M}_{\rm crit}$ is predominantly driven by the accelerated response of the unstable zone width $\Delta R$ beyond this critical point of accretion rate (as shown in Figure \ref{fig:deltaR_mdot}). To further investigate the origin of the rapid $\Delta R$ variations, we find that at the critical accretion rate, the angular momentum transfer at the outer boundary of the unstable zone transitions between two distinct mechanism: from viscous torque dominance (where the disk is hot with small $\beta_{2}$) to magnetic torque dominance (characterized by cooler temperatures and larger $\beta_{2}$). When decreasing the accretion rate from our fiducial value ($\dot{M}=0.1\dot{M}_{\rm Edd}$), the outer boundary of the unstable zone (identified by the intersection between colored and dashed lines in Figure \ref{fig:beta2_R}) displays marked nonlinear response: a $10\%$ reduction in accretion rate ($\dot{M}=0.09\dot{M}_{\rm Edd}$) triggers approximately $50\%$ contraction of the unstable region where magnetic torques govern angular momentum transport, whereas an equivalent accretion rate ($\dot{M}=0.08\dot{M}_{\rm Edd}$) decrease results in merely $\sim 20\%$ additional shrinkage in viscous torque dominate regimes.

\begin{figure}
    \centering
    \includegraphics[width=1\linewidth]{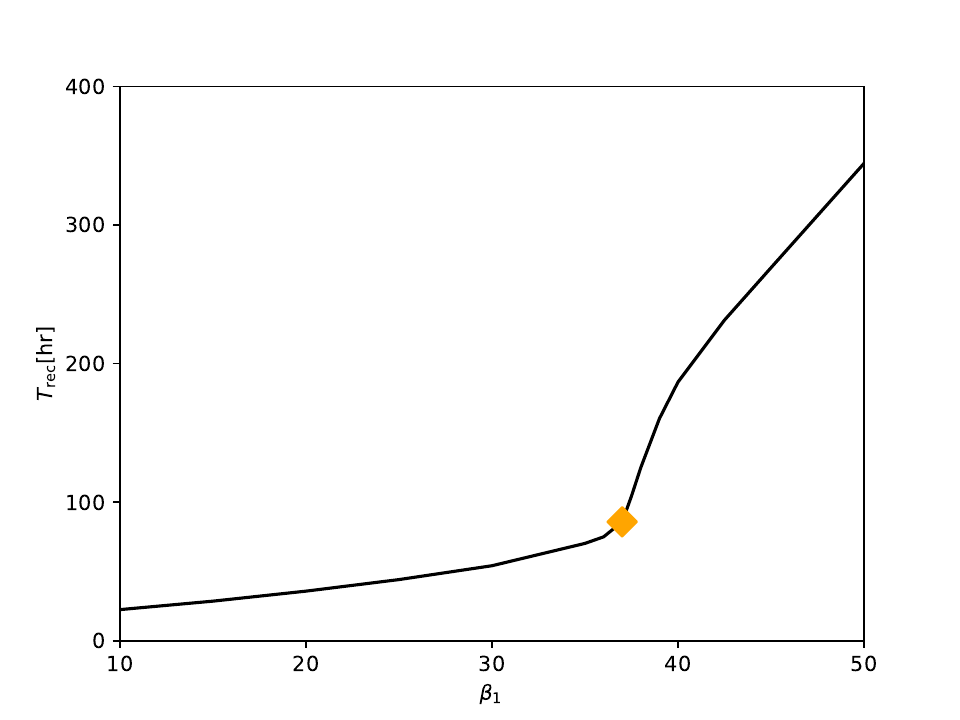}
    \caption{The Recurrence time as a function of magnetic field parameter, where the fiducial parameters are adopted. The yellow diamonds represent critical value $\beta_{1,\rm crit}$ in this relation: when $\beta_{1}<\beta_{1,\rm crit}$ recurrence time shows weak dependence on $\beta_{1}$; when $\beta_{1}>\beta_{1,\rm crit}$ recurrence time become strongly $\beta_{1}$ dependent.}
    \label{fig:rec_beta1}
\end{figure}

The magnetic field parameter $\beta_{1}$ exhibits similar critical behavior, with $\beta_{1,\rm crit}$ demarcating two distinct regimes (Figure \ref{fig:rec_beta1}). Accretion systems with larger magnetic field parameter ($\beta_{1}>\beta_{1,\rm crit}$) display high sensitivity of $T_{\rm rec}$ to $\beta_{1}$ variations, whereas smaller magnetic field parameter disks ($\beta_{1}<\beta_{1,\rm crit}$) maintain relatively stable periods. Importantly, the critical value $\beta_{1,\rm crit}$ measured at fixed $\dot{M}_{0}$ corresponds precisely to the transition point $\dot{M}_{\rm crit}$ ($=\dot{M}_{0}$) obtained in the $T_{\rm rec}-\dot{M}$ relation at fixed $\beta_{1}=\beta_{1,\rm crit}$. This correspondence arises because both transitions reflect the same underlying change in $\beta_{2}$ distribution (Figure \ref{fig:beta2_R}), which indicates whether viscous or magnetic torques dominate angular momentum transport. Consequently, the disk of our model always occupy consistent positions relative to both critical values, either simultaneously in the stable ($\beta_{1}<\beta_{1,\rm crit}$ and $\dot{M}<\dot{M}_{\rm crit}$) or unstable ($\beta_{1}>\beta_{1,\rm crit}$ and $\dot{M}>\dot{M}_{\rm crit}$) regimes.
 
\begin{figure}
    \centering
    \includegraphics[width=1\linewidth]{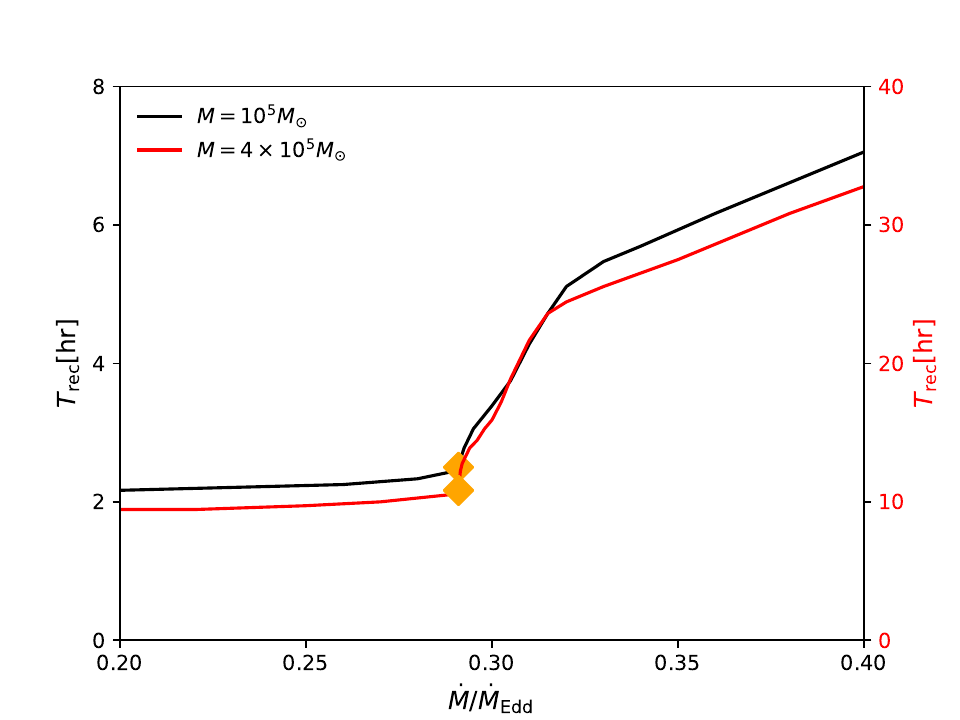}
    \caption{Similar to Figure \ref{fig:rec_mdot}, but using parameter sets corresponding to shorter recurrence times. The black line represents the model with a recurrence time of several hours (similar to eRO-QPE2), adopting the parameters: $M=10^{5}M_{\odot}$, $\beta_{1}=10$, $\alpha=0.1$, $\mu=0.1$, and $f=0.9$. The red line represents the case with a recurrence time of $\sim10$ hours (similar to GSN 069), adopting the similar parameters with black line but $M=4\times10^{5}M_{\odot}$.}
    \label{fig:rec_mdot_sup}
\end{figure}

\begin{figure}
    \centering
    \includegraphics[width=1\linewidth]{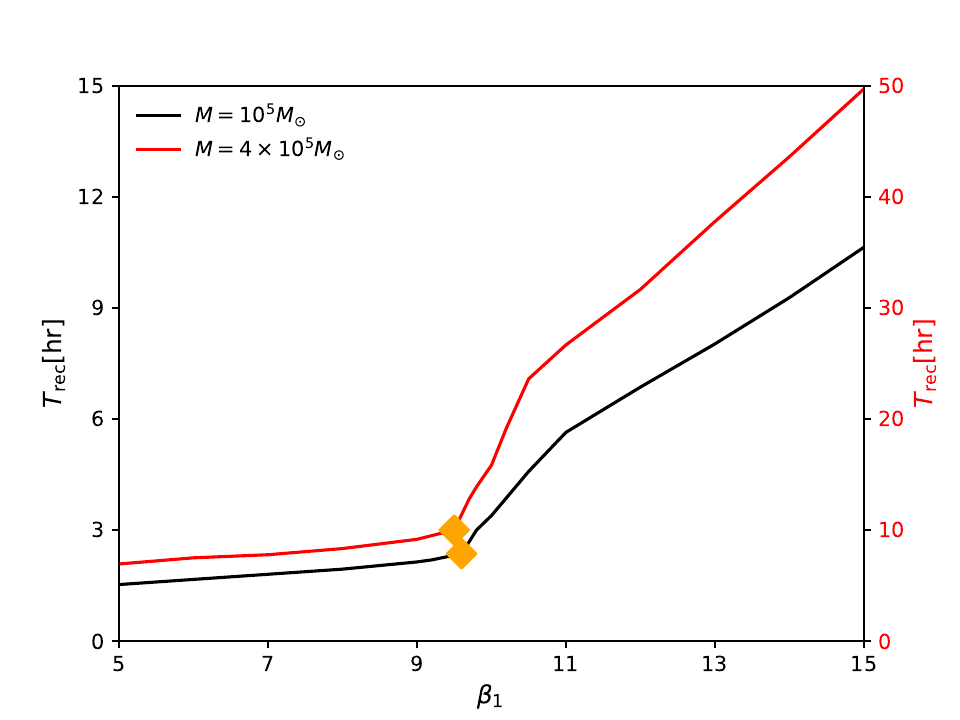}
    \caption{Similar to Figure \ref{fig:rec_beta1}, but using a parameter set corresponding to shorter recurrence times. The black line uses the parameters: $M=10^{5}M_{\odot}$, $\dot{M}=0.3\dot{M}_{\rm Edd}$, $\alpha=0.1$, $\mu=0.1$, and $f=0.9$. The red line uses parameters similar to those of the black line, but with $M=4\times10^{5}M_{\odot}$.}
    \label{fig:rec_beta1_sup}
\end{figure}

 For shorter recurrence times, our fiducial model can achieve $T_{\rm rec}\sim30$ hr with $\beta_{1}=10$ (see Figure \ref{fig:rec_beta1}, where smaller $\beta_{1}$ corresponds to a more centrally concentrated unstable region). To test whether extreme short-period cases like eRO-QPE2 ($T_{\rm rec}\sim2$ hr) and RX J1301 follow the same $T_{\rm rec}-\dot{M}/\beta_{1}$ dependencies found in our fiducial model, we conducted new calculations adopting parameters comparable to those of eRO-QPE2 in \cite{2023ApJ...952...32P}: $M=10^{5}M_{\odot}$, $\dot{M}=0.3\dot{M}_{\rm Edd}$, $\beta_{1}=10$, $\alpha=0.1$, $\mu=0.1$, and $f=0.9$. For systems with periods like GSN 069 ($T_{\rm rec}\sim10$ hr), we used $M=4\times10^{5}M_{\odot}$. As shown in Figures \ref{fig:rec_mdot_sup}-\ref{fig:rec_beta1_sup}, these identical dependencies persist even in short-period regimes.

These results reveal a fundamental dichotomy in QPE behavior controlled by model parameters. In the unstable regime, minor perturbations in either $\dot{M}$ or $\beta_{1}$ induce substantial $T_{\rm rec}$ variations, naturally explaining the highly dispersed periodicity observed in sources like eRO-QPE1, whereas in the stable regime, the system maintains nearly constant $T_{\rm rec}$ values despite significant fluctuations in $\dot{M}$ and $\beta_{1}$, consistent with the regular eruption patterns characteristic of sources like GSN 069.

\subsection{peak temperature vs. $\dot{M}$/$\beta_{1}$}

\begin{figure}
    \centering
    \includegraphics[width=1\linewidth]{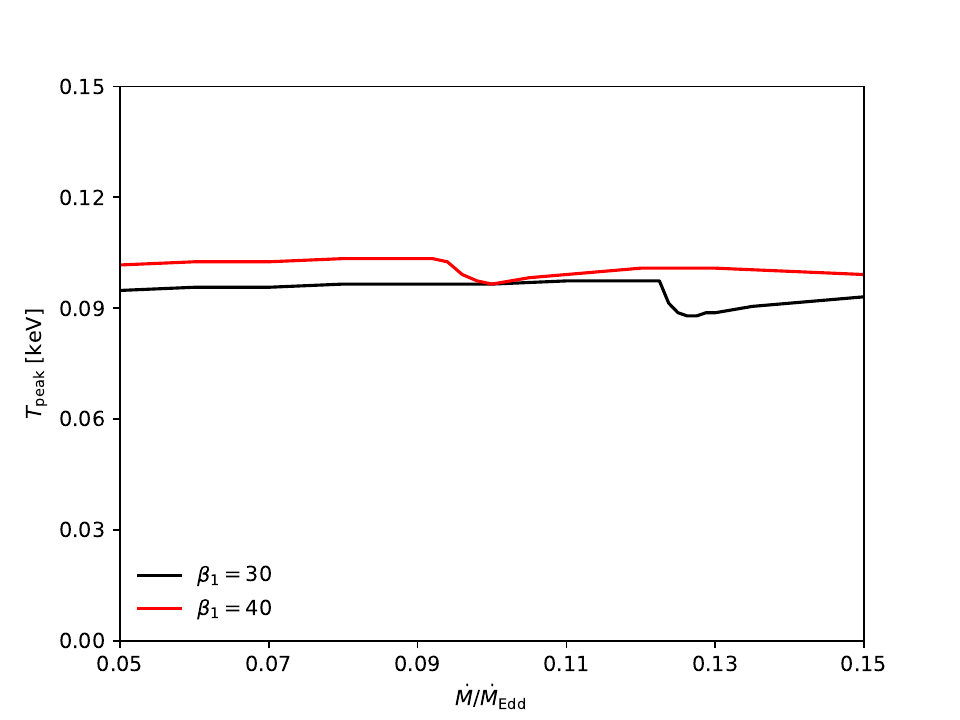}
    \caption{The peak temperature of eruptions as functions of accretion rate. The red line shows the fiducial model with $\beta_{1}=40$, and the black line corresponds to $\beta_{1}=30$.}
    \label{fig:T_peak_mdot}
\end{figure}

\begin{figure}
    \centering
    \includegraphics[width=1\linewidth]{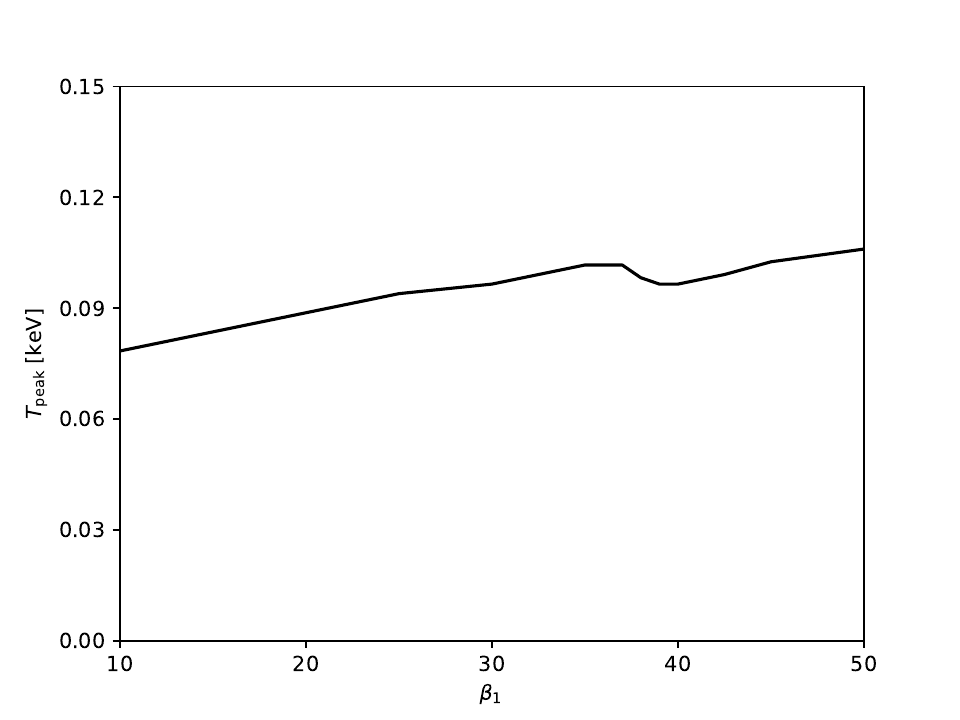}
    \caption{The peak temperature of eruptions as a function of magnetic field parameter, where the fiducial parameters are adopted.}
    \label{fig:T_peak_beta1}
\end{figure}

Similar to the above analysis, we explored the impact of these two parameters on the peak temperature $T_{\rm peak}$ during the burst of limit-cycle in our model by either fixing the magnetic field parameter to change the accretion rate or fixing the accretion rate to change the magnetic field parameter. Our analysis of the peak temperature $T_{\rm peak}$, reveals a fundamental distinction from the repeating period behavior. Unlike the recurrence time $T_{\rm rec}$ which shows critical threshold dependence on accretion rate $\dot{M}$, $T_{\rm peak}$ maintains a stable value across all accretion rate space with no analogous threshold behavior (as shown in Figure \ref{fig:T_peak_mdot}). When analyzing the influence of magnetic field parameter $\beta_{1}$, the peak temperature $T_{\rm peak}$ shows only a weak positive dependence on $\beta_{1}$ (see Figure \ref{fig:T_peak_beta1}), exhibiting fundamentally different behavior from the sharp transition characteristic of the $T_{\rm rec}-\beta_{1}$ correlation. The peak temperature stability arises because outburst temperature are governed mainly by inner boundary physical properties of the unstable zone, not the outer boundary.

This analysis demonstrates that while perturbations in $\dot{M}$ and $\beta_{1}$ may significantly alter eruption periodicity, they exert negligible influence on the peak burst temperature, revealing a key decoupling between the timing and thermal properties of QPEs.

\section{Discussion}

Theoretical explanations for QPEs predominantly follow two distinct physical paradigms: accretion disk instabilities invoking limit-cycle behavior in inner disks and star-disk collision model where a stellar-mass companion periodically impacts the accretion disk of SMBH. While both models can explain certain aspects of QPEs, each faces challenges in accounting for the full range of observed properties. Here, we will discuss these two models by comparing some of their inferences with the observational results of QPEs.

(1): QPEs sources exhibit diverse in recurrence time behaviors, from highly regular (e.g., GSN 069) to strongly stochastic (e.g., eRO-QPE1). The star-disk collision model can account for the regular cases, but struggles to explain irregular eruptions observed in some sources or even in different epochs of otherwise regular QPEs. Our disk instability model with large-scale magnetic field offers a unified explanation: systems with stable recurrence times correspond to the stable parameter regime ($\dot{M}<\dot{M}_{\rm crit}$ and $\beta_{1}<\beta_{1,\rm crit}$, namely left side of the critical thresholds in Figure \ref{fig:rec_mdot} and \ref{fig:rec_beta1}), where perturbations of $\dot{M}$ and $\beta_{1}$ cause minimal period variations. In contrast, sources showing stochastic behavior represent systems in the unstable regime ($\dot{M}>\dot{M}_{\rm crit}$ and $\beta_{1}>\beta_{1,\rm crit}$, namely right side of the critical thresholds in Figure \ref{fig:rec_mdot} and \ref{fig:rec_beta1}), where small parameter fluctuations are amplified into large period changes.

(2): The property of peak temperature during eruptions provides important constraints for theoretical models. The QPEs in GSN 069 exhibit an approximate $L_{\rm bol}\propto T_{\rm bb}^{4}$ correlation, suggesting a constant blackbody emission area \citep{2023A&A...674L...1M}. This qualitatively matches the expectations of star-disk collision model, where the emission area is primarily determined by the radius of the companion star, a parameter that evolves slowly over short periods. However, this relation exhibits significant scatter (particularly for XMM6 and XMM12 in Figure 5 of \citealt{2023A&A...674L...1M}, where large error bars limit the constraint) and is absent in other sources. Notably, eRO-QPE1 maintains nearly constant peak temperatures despite an order-of-magnitude luminosity variation (see Figure 4 in \citealt{2024ApJ...965...12C}), in agreement with our model's prediction that peak temperatures of eruptions remain stable against perturbations in accretion rate and magnetic field parameter (Section 3.2). Furthermore, star-disk collision scenarios lack a natural mechanism to confine the peak temperature of QPEs to the observed $0.1-0.2$ keV range (typical peak temperature range for QPEs).

(3): Several QPE sources exhibit evidence of preceding TDEs, implying a possible physical connection between these phenomena. Theoretical studies have developed a "QPE = TDE + EMRI" scenario \citep{2023ApJ...957...34L,2025ApJ...983L..18J}, where a pre-existing stellar companion interacts with a compact TDE-formed disk to generate QPEs via collisions. This scenario is consistent with some QPE sources with confirmed TDE and their host galaxy properties (recently faded AGN that could retain low-eccentricity stars) \citep{2022A&A...659L...2W,2024ApJ...970L..23W,2025ApJ...983L..18J}. 
However, this scenario faces some challenges, including: the lack of TDE associations in certain QPEs, and their relatively stable peak temperatures during the TDE decay phase (contrary to the $T_{\rm obs}\propto\dot{m}^{11/4}$ relation predicted by \citealt{2023ApJ...957...34L}). These issues suggest the need for further studies on emission mechanisms and QPE production in non-TDE EMRI systems.
In our disk instability framework, the TDE connection emerges naturally through accretion disk evolution. When a TDE occurs around a low-mass SMBH, the extremely high initial accretion rate leads to two distinct phases: 1) during the super-Eddington phase ($\dot{m}\gtrsim1$), advection dominates the balance of energy equation, potentially suppressing instability entirely, while 2) as the accretion rate declines to sub-Eddington levels ($\dot{m}<1$), an unstable region develops and induce accretion disk variability. Radiation pressure instability grows in disk regions where radiation pressure dominates the total pressure (see Section 4 of \citealt{2022ApJ...928L..18P} and Equation 4 in \citealt{2023MNRAS.524.1269K}), provided the viscous torque depends primarily on total pressure ($\mu<0.56$). Short-timescale repeating eruptions only appear when the accretion rate decreases to an appropriate threshold value (coupled with sufficient magnetic field strength), allowing magnetic confinement of the unstable region to a limited radial extent (analogous to Figure \ref{fig:deltaR_mdot}). However, rapid declines in the accretion rate prevent sustained eruptive activity, and overly violent perturbations in some sources degrade signal coherence, making periodicity detectable only through Lomb-Scargle periodograms rather than via direct light curve inspection. While our model demonstrates that radiation pressure instability can produce eruptions when the accretion rate falls below the Eddington limit ($\dot{M}<\dot{M}_{\rm Edd}$), the exact timing of QPE detection post-TDE remains quantitatively uncertain due to complex parameter dependencies and observational constraints. However, our framework consistently requires low accretion rates ($\dot{m}\sim0.1$) for short-period QPE production. This condition is reached earlier in systems with either shorter viscous timescales \citep{2025arXiv250420148G} or larger BH masses \citep{2012ApJ...760..103D}. We suggest this explains why AT2019vcb exhibited QPEs sooner after TDE than GSN 069 or AT2019qiz, as it likely had more rapid accretion rate evolution and/or higher BH mass. For TDEs with sufficiently weak magnetic fields, even at low accretion rates ($\dot{m}\sim0.01$) the unstable region's outer boundary remains extended. Consequently, such weakly magnetized systems cannot produce short-timescale QPE-like eruptions, but may instead exhibit longer-term variability observable in lower energy bands (e.g., UV/optical).

As an example of applying our model to the long-term evolution of QPEs, we use our model to interpret two representative sources: GSN 069 and eRO-QPE1. Following its 2010 TDE, GSN 069 likely experienced initial high accretion with suppressed or long timescale variability during the early decay phase. As the accretion rate decreased, radiation pressure instability developed and became constrained to shorter timescales, producing the observed $\sim9$ hr QPEs. The source then entered the stable parameter regime, maintaining relatively constant periods until its 2020 rebrightening (likely a partial TDE; \citealt{2023A&A...670A..93M, 2025arXiv250420148G}). This event temporarily increased $\dot{M}$ beyond the stable range, suppressing detectable QPEs until the accretion rate decreased sufficiently to restore short period eruptions. Our model shows that when both the accretion rate and magnetic field parameter simultaneously increase by $30\%$ from the parameter set corresponding to the yellow diamond on the red line of Figure \ref{fig:rec_mdot_sup} (which produces QPE periods similar to GSN 069's observed $\sim9$ hr cycle), the QPE period increases by $\sim360\%$. This extends the recurrence time from $\sim9$ hr to $\sim41$ hr ($\sim150$ ks), longer than $\sim120$ ks observational windows. Such co-evolution of $\dot{M}$ and $\beta_{1}$ during long-term TDE evolution is physically plausible, potentially explaining why QPEs were not detected in GSN 069 in 2014 but were observed in 2018. The observed period stability during 2018-2020 despite moderate $\dot{M}$ variations implies moderately small $\beta_{1}$ values in GSN 069, permitting stability across a broad range of accretion rates. During GSN 069's rebrightening phase, the observed decreases in both temperature and outburst intensity (Figure 2 of \citealt{2023A&A...670A..93M}) may be explained by our model as resulting from a gradual increase in magnetic field strength, a trend consistent with earlier QPE evolution. This behavior could arise from delayed propagation of the enhanced accretion rate to the innermost region, where the X-ray luminosity from the stable region outside the unstable zone increases first while the full impact on QPE properties manifests only after the increased matter reaches the ISCO. In contrast, eRO-QPE1 displays persistently unstable periods, with particularly erratic behavior emerging after October 2022 despite an overall accretion rate declining $\dot{M}$ trend \citep{2024ApJ...965...12C}. Its low quiescent luminosity may originate from either extremely small $\beta_{1}$ values that facilitate efficient outflow driven energy loss, or intrinsically low accretion rate. Our Figure \ref{fig:rec_mdot} indicates eRO-QPE1 most likely occupies the high $\beta_{1}$, low $\dot{M}$ parameter space where QPE periods remain highly sensitive to perturbations across nearly all accretion rates.

Our model exhibits a notable limitation in its predicted correlation between eruption properties. Since the eruption amplitude is intrinsically tied to the size of the unstable region ($\Delta R$) while the peak temperature remains stable, the model naturally produces a positive correlation between eruption amplitude and recurrence time. This prediction shows reasonable agreement with observational data from sources like GSN 069 and eRO-QPE2 (see Figure 1 of \citealt{2022A&A...662A..49A}), where larger amplitude eruptions tend to have longer recurrence times. However, clear discrepancies emerge when examining eruptions in RX J1301 \citep{2024A&A...692A..15G} and eRO-QPE1 \citep{2024ApJ...965...12C}, where at least a subset of eruptions have no such correlation. More strikingly, some cases like eRO-QPE1 obs1 \citep{2022A&A...662A..49A} even exhibit complex light curves (possibly arising from gravitational lensing effects; \citealt{2025arXiv250605517X}). These discrepancies may originate from two mechanisms: 1) In low accretion rate environments, the innermost region of disk may undergo state transitions between thin-disk and advection-dominated accretion flow (ADAF) configurations; 2) For accretion systems in the unstable parameter regime (right side of Figures \ref{fig:rec_mdot} and \ref{fig:rec_beta1}), strong radiative feedback during eruptions may substantially alter instability growth before complete matter replenishment reaches the ISCO (we will further investigate this in the future). 

One discrepancy between our model and observations concerns the light curve profiles: while QPE observations typically show rapid rises followed by slow decays, our model's light curves exhibit the opposite behavior (see \citealt{2022ApJ...928L..18P,2023ApJ...952...32P}). This rapid decay in our model primarily originates from the one-zone treatment of the instability. When eruption occurs, material in the unstable region is rapidly accreted by the black hole, depleting the inner disk region that produces high-temperature radiation. However, in reality, after an outburst driven by radiation pressure instability, a portion of the inner disk material would be ejected in the vertical direction rather than being immediately accreted. This vertical spread of matter would sustain high-temperature radiation for longer durations, resulting in a slower decay than predicted by our toy model. On the other hand, as mentioned in the Introduction, QPEs generally exhibit counter-clockwise hysteresis cycles in the L-T plane. This hysteresis behavior physically represents the initial heating followed by cooling with expansion maintained throughout QPE eruptions, a pattern fully consistent with the expected radiation pressure instability scenario in accretion disks. However, our current model adopts a simplified treatment where the width of the unstable zone, $\Delta R$, remains fixed throughout eruptions. This fixed $\Delta R$ approximation prevents our model from reproducing the observed hysteresis loops in the L-T plane. Future investigations with more complete disk models will explore this important observational feature in detail.

\section{Conclusion}
In this work, we have systematically investigated how accretion rate ($\dot{M}$) and magnetic field parameter ($\beta_{1}$) perturbations affect the properties of QPEs within the framework of a disk instability model with large-scale magnetic field. We identify critical thresholds ($\dot{M}_{\rm crit}$, $\beta_{1,\rm crit}$) that divide recurrence time of model into two distinct regimes:
\begin{itemize}
\item A stable regime ($\dot{M}<\dot{M}_{\rm crit}$, $\beta_{1}<\beta_{1,\rm crit}$) maintains nearly constant eruption periods against perturbations in both accretion rate and magnetic field parameter;
\item An unstable regime ($\dot{M}>\dot{M}_{\rm crit}$, $\beta_{1}>\beta_{1,\rm crit}$) where perturbations in these parameters generate significant period variations.
\end{itemize}

\noindent While recurrence times show strong dependence on both accretion rate and magnetic field parameter:
\begin{itemize}
\item The peak temperatures of QPEs exhibit relative stability despite variations in the accretion rate and magnetic field parameter.
\end{itemize}
This framework naturally accounts for both regular (e.g., GSN 069) and stochastic (e.g., eRO-QPE1) eruption QPEs, as well as the observed small scatter in QPE peak temperatures during long-term evolution.

Future studies may identify more QPEs in archival data from XMM-Newton and eROSITA, as well as through ongoing X-ray surveys like Einstein Probe \citep{2025arXiv250107362Y}. These expanded samples will enable statistically robust examinations of the TDE-QPE connection through their association rates, while also providing tighter constraints on the distribution of eruption peak temperatures and potentially revealing new observational features. Such findings would critically test and refine theoretical models.

\begin{acknowledgments}
We thank the referee for the helpful comments which helped to improve various aspects of this paper. XP thanks Dr. Huaqing Cheng for useful comments. This work is supported by the Postdoctoral Fellowship Program of CPSF (Grant NO. GZC20232696), National Natural Science Foundation of China (Grant NO. 12433005, 12333004, 12273089, 12073023, 12233007, 12361131579, and 12347103), the science research grants from the China Manned Space Project with No. CMS-CSST-2021-A06, and the fundamental research fund for Chinese central universities (Zhejiang University).
\end{acknowledgments}


\bibliography{paper}{}
\bibliographystyle{aasjournalv7}



\appendix
\section{dependence between $\dot{M}$ and $\beta_{1}$ perturbations}
The modeled variations in recurrence time and peak temperature with respect to accretion rate and magnetic field parameter presented in previous sections were derived by independently varying either parameter while holding the other constant. However, in realistic accretion disks, matter and magnetic fields are fundamentally coupled. Changes in the magnetic field can alter the radial inflow velocity of disk matter, which may consequently modify the local accretion rate. Conversely, variations in accretion rate may affect the total-to-magnetic pressure ratio that determines $\beta_{1}$. Here, we perform a simple qualitative analysis to examine the coupling between perturbations in $\dot{M}$ and $\beta_{1}$.

The radial velocity of accreting matter can be derived through integration of the angular momentum equation (Eq. \ref{angularmom}) in combination with the mass continuity relation $\dot{M}=2\pi R\Sigma V_{\rm R}$, yielding:
\begin{equation}
    V_{\rm R}=-\frac{R\mathscr{B}\mathscr{C}^{-1/2}\mathscr{D}T_{r\phi}}{\Sigma\left(l_{\rm k}-\frac{f\dot{M}_{\rm Rin}}{\dot{M}}l_{\rm k,Rin}\right)} + \frac{\int_{\rm R_{\rm in}}^{\rm R} RT_{\rm m}{\rm d}R}{R\Sigma\left(l_{\rm k}-\frac{f\dot{M}_{\rm Rin}}{\dot{M}}l_{\rm k,Rin}\right)},
    \label{integ_angularmom}
\end{equation}
where $R_{\rm in}$ is the radial position of ISCO. The first term on the right-hand side of Equation (\ref{integ_angularmom}) represents the radial velocity component due to viscous torque, $V_{\rm R,vis}$, while the second term represents the velocity component induced by magnetic torque, $V_{\rm R,m}$.

For this qualitative analysis, we implement a simplified treatment that neglects general relativistic corrections, and extracts the magnetic torque term $T_{\rm m}$ from the integral by assuming a power-law radial dependence $T_{\rm m}\propto R^{n}$. This yields:
\begin{equation}
    V_{\rm R,vis}\sim -\frac{RT_{\rm r\varphi}}{\Sigma\left(l_{\rm k}-\frac{f\dot{M}_{\rm Rin}}{\dot{M}}l_{\rm k,Rin}\right)},
    \label{VRvis}
\end{equation}
\begin{equation}
    V_{\rm R,m}\sim \frac{RT_{\rm m}}{\Sigma\left(l_{\rm k}-\frac{f\dot{M}_{\rm Rin}}{\dot{M}}l_{\rm k,Rin}\right)}.
    \label{VRm}
\end{equation}
General form of viscous torque $T_{\rm r\varphi}=-2\alpha P^{1-\mu}P_{\rm gas}^{\mu}H$ and magnetic torque $T_{\rm m}=C_{0}PR$ ($C_{0}$ is a factor decided by magnetic field configuration and in this work $C_{0}=0.508/(1+\beta_{1})$) can be incorporated into the above equations. This allows us to derive the ratio between the corresponding velocity components:
\begin{equation}
    \frac{V_{\rm R,vis}}{V_{\rm R,m}}=\frac{2\alpha P^{1-\mu}P_{\rm gas}^{\mu}H(1+\beta_{1})}{0.508PR}
    \label{VR_ratio}.
\end{equation}
For parameter values where $\mu$ remains moderate and $\beta_{1}\gg1$, the velocity ratio scales as $V_{\rm R,vis}/V_{\rm R,m}\sim \beta_{1}H/R$ , being primarily determined by the disk aspect ratio and magnetic field parameter. When magnetic fields are non-negligible, this ratio becomes $\ll1$, indicating magnetic torque dominance in angular momentum transport.

\begin{figure}
    \centering
    \includegraphics[width=0.7\linewidth]{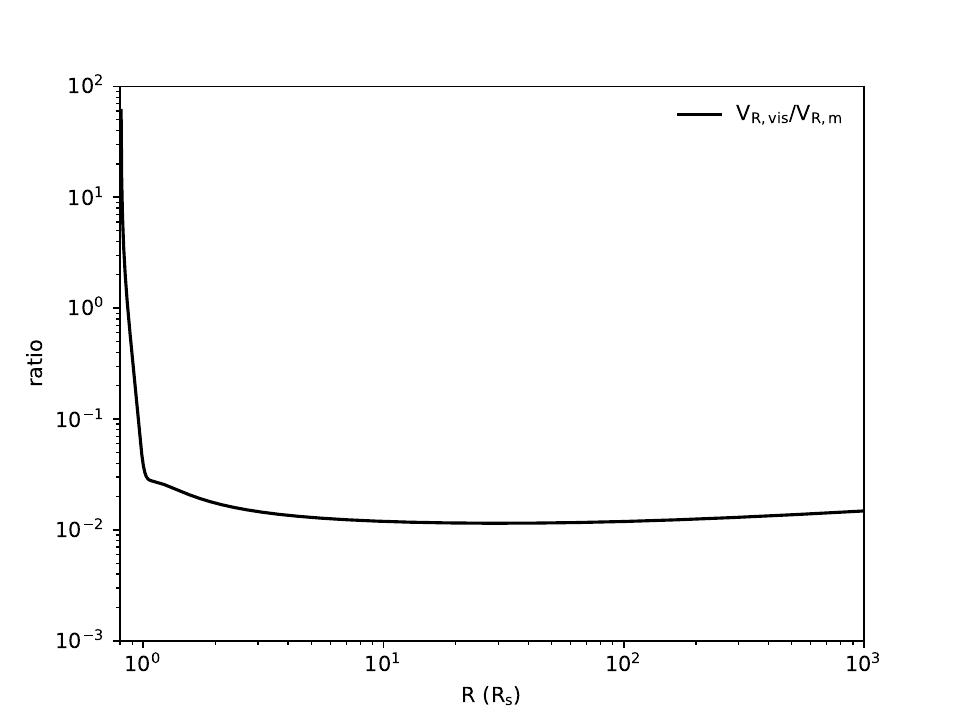}
    \caption{Numerically calculated ratio of radial velocities driven by viscous torque to magnetic torque as a function of radius, obtained using the model described in Section 2 with fiducial parameters.}
    \label{fig:velocity_ratio}
\end{figure}

As a comparison, Figure \ref{fig:velocity_ratio} presents the numerically computed velocity ratio derived from Section 2. The results confirm that this ratio remains $\ll1$ across the majority of the disk, with the exception of the innermost radiation-pressure-dominated region where the flow transitions to an unstable state.

Therefore, based on the above derivation and incorporating the continuity equation, we can establish the relationship between the local accretion rate and magnetic field parameter:
\begin{equation}
    \dot{M}\approx2\pi R\Sigma V_{\rm R,m}\propto T_{\rm m}\sim\frac{P}{\beta_{1}}.
    \label{acc_rate}
\end{equation}

Furthermore, by including both the viscous torque $T_{\rm r\varphi}$ and the electron-scattering optical depth $\tau=\kappa_{\rm es}\Sigma/2$ in Equation (\ref{energy}), we derive an approximate pressure-temperature relation under the condition of moderately small $\mu$:
\begin{equation}
    P\sim (\frac{8ac\Omega_{\rm k}\mathscr{C}}{9\alpha\kappa_{\rm es}\mathscr{B}\mathscr{D}})^{1/2}T_{\rm c}^{2},
    \label{P-T}
\end{equation}
where vertical hydrostatic equilibrium $H=c_{\rm s}/\Omega_{\rm k}$ is adopted.

Here, we consider perturbations of three fundamental disk quantities: magnetic pressure $P_{\rm m}$ ($=B^{2}/8\pi$), temperature of midplane $T_{\rm c}$, and surface density $\Sigma$. Derived quantities like total pressure $P$ and scale height $H$ can be expressed as functions of these primary variables. We then assess whether the induced perturbations in both the magnetic field parameter $\beta_{1}$ and accretion rate $\dot{M}$ can be treated as independent variables.

A local perturbation in magnetic pressure $\delta P_{\rm m}$ generates: (1) a corresponding perturbation in magnetic torque $\delta T_{\rm m}$ (where $T_{\rm m}=B_{\rm z}B_{\varphi}R/2\pi$), which produces an accretion rate variation $\delta\dot{M}/\dot{M}=\delta P_{\rm m}/P_{\rm m}$; and (2) an inverse variation in the magnetic field parameter $\delta\beta_{1}/\beta_{1}=-\delta P_{\rm m}/P_{\rm m}$, since the total pressure $P$ is primarily governed by $\Sigma$ and $T_{\rm c}$. This reveals the $\beta_{1}$-$\dot{M}$ coupling through $\delta P_{\rm m}$. However, the width of unstable region $\Delta R$, determined locally by $\beta_{2}$ ($\sim P_{\rm gas}/P$), remains unaffected by $\delta P_{\rm m}$. Consequently, the instability recurrence time $T_{\rm rec}\sim \Delta R/V_{\rm R}$ exhibits linear scaling $\delta T_{\rm rec}/T_{\rm rec}=-\delta P_{\rm m}/P_{\rm m}$. This demonstrates that $\dot{M}$, $\beta_{1}$, and $T_{\rm rec}$ respond equally to magnetic pressure perturbations.

A temperature perturbation $\delta T_{\rm c}$ induces a variation in the magnetic field parameter $\delta\beta_{1}/\beta_{1}= \delta(P/P_{\rm m})/(P/P_{\rm m})= 2\delta T_{\rm c}/T_{\rm c}$ (ref Equation \ref{P-T}), while leaving the accretion rate unchanged (ref Equation \ref{acc_rate}). This demonstrates that $\beta_{1}$ perturbation caused by $\delta T_{\rm c}$ can be analyzed independently of $\dot{M}$ perturbation. 

Under magnetic flux freezing (perfect matter-field coupling), a surface density perturbation $\delta\Sigma$ produces magnetic pressure variations $\delta P_{\rm m}/P_{\rm m}=2\delta\Sigma/\Sigma$. In gas-pressure-dominated regions, the total pressure responds as $\delta P/P=4\delta\Sigma/3\Sigma$, as derived from Equation (\ref{P-T}) combined with vertical hydrostatic equilibrium considerations. This pressure response results in an accretion rate perturbation $\delta\dot{M}/\dot{M}=2\delta\Sigma/\Sigma$ that is three times stronger than the corresponding magnetic parameter variation $\delta\beta_{1}/\beta_{1}=-2\delta\Sigma/3\Sigma$. Therefore, perturbations in both $\beta_{1}$ and $\dot{M}$ induced by $\Sigma$ variations can be treated as approximately independent variables.

Consequently, perturbations in the accretion rate and magnetic field parameter exhibit both coupled and independent conditions. However, in the coupled case, the amplitude of period dispersion scales linearly with $\dot{M}/\beta_{1}$ perturbation amplitudes ($\delta\dot{M}/\dot{M}\sim\delta\beta_{1}/\beta_{1}$ is small), resulting in negligible period modulation. We therefore focus primarily on scenarios where these parameters vary independently, which generate more significant observational signatures.

\end{document}